\documentclass[10pt,preprint]{aastex}
\usepackage{natbib} % use author/date bibliographic citations

\usepackage{amsmath,amssymb}%, mathtools}  % Better maths support & more symbols
\usepackage{bm}  % Define \bm{} to use bold math fonts
\usepackage{array} % for better arrays (eg matrices) in maths
\usepackage{multirow}
\usepackage{slashbox}

\usepackage{textcomp} % provide lots of new symbols
\usepackage{ulem}
\usepackage{verbatim} % adds environment for commenting out blocks of text & for better verbatim
\usepackage[usenames,dvipsnames]{color}
\usepackage{indentfirst}
\usepackage{titlesec, sectsty}
\usepackage{graphics,graphicx} % Add graphics capabilities
\usepackage{subfigure} % make it possible to include more than one captioned figure/table in a single float
\usepackage{wrapfig}
\usepackage{rotating}

\usepackage[top=1in,bottom=1in,left=1in,right=1in]{geometry} % to change the page dimensions
\sectionfont{\normalsize}
\subsectionfont{\normalsize}
\subsubsectionfont{\normalsize}
\renewcommand{\it}{\itshape}

\renewcommand{\rm}{\mathrm}
\newcommand{\msol}[1]{\mbox{#1~M$_{\odot}$}}

\newcommand{\rsol}[1]{\mbox{#1~R$_{\odot}$}}

\newcommand{\gcmcu}[1]{\mbox{#1~g~cm$^{-3}$}}
\newcommand{\kms}[1]{\mbox{#1~km~s$^{-1}$}}

\newcommand{\Ni}{\mbox{$^{56}$Ni}}
\newcommand{\Fe}{\mbox{$^{56}$Fe}}

\begin{document}

\title{First Simulations of Core- Collapse Supernovae to Supernova Remnants with SNSPH}
\author{C. I. Ellinger}
  \affil{The University of Texas at Arlington}
  \affil{Department of Physics, 502 Yates Street, Arlington, TX 76019, USA}
  \email{carola.ellinger@gmail.com}
\author{G. Rockefeller, C. L. Fryer\altaffilmark{1,2}}
  \affil{Los Alamos National Laboratory}
  \affil{CCS Division, Los Alamos, NM, USA}
\author{P. A. Young}
  \affil{Arizona State University}
  \affil{School of Earth and Space Exploration, Tempe, AZ 85287, USA}
\author{S. Park}
  \affil{The University of Texas at Arlington}
  \affil{Department of Physics, 502 Yates Street, Arlington, TX 76019, USA}
  
\altaffiltext{1}{The University of Arizona, Department of Physics, Tuscon, AZ 85721, USA}
\altaffiltext{2}{The University of New Mexico, Department of Physics, Albuquerque, NM 87131, USA}

\shorttitle{First SN to SNR simulations in 3D}
\shortauthors{Ellinger et al.}


\date{} % delete this line to display the current date

%%%%%%% FORMATTING

%\pagestyle{plain}
%\pagenumbering{arabic}

%\maketitle %make title for article class
%\tableofcontents

%~~~~~~~~~~~~~~~~~~~~~~~~~~~~~~~~~~~~~~~~~~~~~~~~~~~~~~~~~~~~
%%%%%%%%%%%%%%%%   ABSTRACT   %%%%%%%%%%%%%%%%
%~~~~~~~~~~~~~~~~~~~~~~~~~~~~~~~~~~~~~~~~~~~~~~~~~~~~~~~~~~~~

\begin{abstract}{\small
We present the first 3-dimensional simulations following the evolution of supernova 
shocks from their inception in the stellar core through the development of a supernova 
remnant into the Sedov phase.  Our set of simulations use two different progenitors and two 
different conditions for the structure of the circumstellar environment.  These calculations 
demonstrate the role that supernova instabilities (the instabilities that develop as 
the shock drive through the star) play in defining the structure and long-term 
development of instabilities in supernova remnants.  We also present a first 
investigation of the mixing between stellar and interstellar matter as the supernova 
evolves into a young supernova remnant.
}
\end{abstract}

\maketitle %make title for aastex class

%{\small
%\twocolumn
%~~~~~~~~~~~~~~~~~~~~~~~~~~~~~~~~~~~~~~~~~~~~~~~~~~~~~~~~~~~~
%%%%%%%%%%%%%%%%   SECTION 1   %%%%%%%%%%%%%%%%
%~~~~~~~~~~~~~~~~~~~~~~~~~~~~~~~~~~~~~~~~~~~~~~~~~~~~~~~~~~~~
\section{Introduction\label{s:intro}}

Supernova remnants (SNR) provide a window into the inner workings of the
supernova engine by providing insight into both the nucleosynthetic
yields and the asymmetries that occur in the explosion.  But numerical
models are required to connect observations of the remnant to the
exact behavior of the supernova engine.  As such, these models
facilitate both improved interpretations of the observations as well
as improve our understanding of the explosion mechanism.  In the past,
the modeling of core-collapse supernovae (CC-SNe) in 3 dimensions was
usually aimed at either the explosion of the star
\citep[e.g.][]{HungerfordFW03, HungerfordFR05, HammerJM10,JoggerstAW10, eyfr11} or the
evolution as SNR \citep[e.g.][]{Fraschetti_ea10,
  vMarle10, Orlando_ea12,WarrenB13}. This, however, ignores the intimate link
that exists between the SNR morphology and turbulent phases in the SN
explosion, and the potential for using the observational signatures of
the large scale mixing of heavy elements to derive constraints on the
progenitor system and explosion dynamics.  It has been recognized for
a long time that the evolution of a SNR becomes turbulent already at
the earliest evolutionary stages \citep{Gull73a, ChevalierBE92}.
Starting with the explosion of the massive star itself, even in
spherically symmetric conditions, multiple episodes of extensive,
Rayleigh-Taylor (RT) instability-induced mixing break the symmetry of
the expanding material \citep[e.g.][]{HammerJM10,eyfr11,Fraschetti_ea10,Orlando_ea12}.

In a CC-SN explosion, the chemical properties of the ejecta are intimately linked to the structure 
of the progenitor star and dynamics of the explosion.  Nucleosynthesis calculations in 1D have 
provided yields for an array of parameterized, spherically symmetric SN explosions 
\citep[e.g.][]{WW95,Rauscher_ea02}. These models are very unconstrained \citep{YoungF07} 
though somewhat useful for cases where the total yields of the synthesized elements 
in a SNR are known. But they are difficult to apply to resolved individual structures 
in SNRs where, as is evident from observational studies \citep[e.g.][]{Hwang_ea04, HwangL12, Isensee_ea10} turbulent flows have resulted in partial mixing of material across the layers.

Young SNRs form by the ejecta sweeping up the surrounding material and forming a reverse shock that propagates inwards and heats the gas to very high (ionizing) temperatures \citep[e.g.][]{TrueloveM99,Chevalier82}. 
Various phases of this interaction can be observed via the emission it produces, e.g. X-rays trace the shocked material in high ionization states, optical emission detects low ionization states, infrared traces emission from cold, unshocked ejecta and dust, radio waves are emitted by synchrotron electrons in the pulsar wind nebula and at the shock front, and gamma rays can potentially detect the decay lines of $^{56}$Ni and $^{44}$Ti produced in the explosion, or cosmic rays accelerated at the SN shock front \citep[e.g.][]{EriksenPHD,Martin_ea09,DermerP13,FanLF10}. 

In general, the spatially resolved abundance ratios in SNRs can be used to differentiate between 
material processed in the star's hydrostatic burning lifetime and that processed explosively 
\citep{arnett_1996_aa}. The degree of explosive processing combined with the kinematics of the ejecta 
can be interpreted to find the energy of the explosion and identify the locations in the star and 
explosion where the asymmetries arose. One may be able to distinguish between asymmetries seeded by 
pre-collapse hydrodynamic convection \citep{MeakinA06,Ott_ea13,Fryer04,BurrowsH96}, convection in the neutrino gain region 
during the explosion \citep[e.g.][]{Mezzacappa_ea98conv,FryerHR07,BurrowsHF95,Herant_ea94}, and global mode asymmetries seeded by standing accretion 
shock instabilities (SASI) \citep[e.g.][]{BlondinMD03,Hanke_ea12}, rotation, or jets from the central engine. 

High resolution observations of young SNRs made possible by NASA's {\it Great Observatories} have 
demonstrated the existence of turbulent motions during and after the explosion at all 
resolvable length scales \citep[e.g. ejecta knots of various elements in Cassiopeia A (Cas A) and G292.0+1.8 (G292), Ni in 
SN 1987A][]{Park_ea02, Park_ea07, Hwang_ea04}.
As evident from X-ray observations, which trace mainly the ejecta distribution behind the 
reverse shock, this material consists of a vast network of filaments and fast moving knots
showing bright emission lines \citep[e.g.][]{Park_ea02, Park_ea07, Hwang_ea04}. 
Analysis of the spectra of filamentary features in G292 and Cas A shows that a majority of this material is metal-rich ejecta \citep[e.g.][]{HwangL12, Park_ea07, Bhalerao_ea13} which, as theory suggests, has begun to mix with the surrounding medium. Numerical simulations of the SNR evolution let us study what processes acted on an individual parcel of gas to place it where it is observed within the SNR, and would let us untangle the history of fluid instabilities that influenced the dynamical paths of different parcels of gas. 

Most observable young SNRs have progressed to a stage where the interaction with the ambient environment has become important.
Cas A, e.g., has been estimated to have swept- up about $2-3$ times its ejecta mass in surrounding material \citep{VinkKB96}, and G292 has been estimated to have swept- up $1-2$ times its ejecta mass in surrounding material \citep{Lee_ea10}. At this stage, some deceleration has occurred in the outermost ejecta, and mixing between ejecta and circumstellar material is becoming evident.

In this paper we present the first numerical calculations in 3D of CC-SNe followed (nearly) seamlessly into the Sedov-Taylor (ST) stage of SNRs, or until the reverse shock is well on its way back to the center of the explosion. We limit our study to two progenitors exploding as spherically symmetric CC-SNe and expanding into surrounding media with simple density structures, two of constant density and one of powerlaw profile. The setup of these initial conditions and the simulations 
are discussed in detail in Section \ref{s:sims}. The instabilities that occur in the remnant evolution are presented in Section \ref{s:results} and discussed in Section \ref{s:conclude}.

%\input{./background.tex}

%~~~~~~~~~~~~~~~~~~~~~~~~~~~~~~~~~~~~~~~~~~~~~~~~~~~~~~~~~~~~
%%%%%%%%%%%%%%%%   SECTION 2   %%%%%%%%%%%%%%%%
%~~~~~~~~~~~~~~~~~~~~~~~~~~~~~~~~~~~~~~~~~~~~~~~~~~~~~~~~~~~~
\section{Simulations\label{s:sims}}

\subsection{The Progenitors and Explosions\label{s:expl}}

For the initial conditions we use two different stellar
progenitors: a solar metallicity, \msol{20} red supergiant (star) RSG
and the binary \msol{16} progenitor from \citet{Young_ea06}. In part, these progenitors were
chosen to resemble the progenitors that have been inferred for G292
\citep{Lee_ea09} and Cas A \citep{Young_ea06}.  In follow- up papers, we
will study a full suite of SNR models for comparison to these remnants.
We mimicked a binary common envelope phase in the \msol{16} star by 
removing its hydrogen envelope when the star is at the base of the
first- ascent red giant branch.  Both stars were evolved up to the onset of
core collapse with the stellar evolution code TYCHO \citep{YoungA05}.
The models are non-rotating and include the hydrodynamic mixing
processes described in \cite{YoungA05, Young_ea05, ArnettMY09, ArnettMY10, ArnettM11}. The
inclusion of these processes, which approximate the integrated effect
of dynamic stability criteria for convection, entrainment at
convective boundaries, and wave-driven mixing, results in
significantly larger extents of regions processed by nuclear burning
stages.  Mass loss uses updated versions of the prescriptions of
\citet{Kudritzki_ea89} for OB mass loss and \citet{Bloecker95} for red
supergiant mass loss, and \citet{LamersN03} for WR phases.  A 177
element network terminating at $^{74}$Ge is used throughout the
evolution. The network uses the most current Reaclib rates
\citep{RauscherT01}, weak rates from \citet{LangangkeM00}, and
screening from \citet{Graboske_ea73}. Neutrino cooling from plasma
processes and the Urca process is included.

To model collapse and explosion, we use a 1- dimensional Lagrangian
code to follow the collapse through core bounce.  This code includes
3-flavor neutrino transport using a flux-limited diffusion calculation
and a coupled set of equations of state to model the wide range of
densities in the collapse phase \citep[see][for
  details]{Herant_ea94, Fryer99}. It includes a 14-element nuclear
network \citep{BenzTH89} to follow the energy generation.  Following
the beginning of the explosion in 1D saves computation time and is
sufficient for this problem, as we were mainly interested in the
formation of structure during the passage of the shock.  The explosion
was followed until the revival of the shock, and then mapped into 3D
to follow the rest of the explosion and further evolution in 3
dimensions. The mapping into an optimized, 3D distribution of SPH
particles was accomplished using the WVT algorithm described in \citep{wvt}.
The mapping took place when the supernova shock wave has moved out of
the Fe-core and propagated into the Si-S rich shell, i.e. shortly
after the revival of the bounce- shock.

\subsection{Model Interstellar Media\label{ss:isms}}

We consider two different ISM scenarios of constant density in this study, one presenting a cold molecular cloud (MC) type environment (using $\rho$ \gcmcu{$\approx 2.4\times10^{-19}$}, $T = 10~K$), and one a cold neutral interstellar medium (CNM) type (using $\rho$ \gcmcu{$\approx 1.4\times10^{-22}$}, $T = 80~K$). 
Explosions into the molecular cloud or the cold neutral medium will present the limiting cases where the progenitor did not leave a significant imprint on its environment. Since the constant density ISM was assumed to extend down to the surface of the star in each case, the ages derived from each simulation should not be taken as a direct correspondence to physical ages of SNRs. Most SNe would sweep through an extensive region of low- density gas before interacting with a higher density ISM like a cold neutral ISM or a molecular cloud. The ejecta would have to traverse a greater distance in order to sweep up enough material to interact with it in a lower density ISM. To first order, the density of the (constant) ISM determines the time scale of the interaction and the age of the remnant at particular stages, thus a better indicator for the age of the simulated remnant is the value of the mass ratio m$_{ejecta}$/m$_{swept-up}$. There are not expected to be vast morphological differences between different (constant) density ISMs. The main difference will be in the age of a SNR for different density ISMs at the same evolutionary stage. 
We run the SN explosions out until the ejecta has swept up a mass of surrounding material equal to ten times its own mass. 
This puts the SNR models at a slightly more evolved stage than Cas A and G292. 
At this point the young, or ejecta dominated SNR phase is nearing the transition to the snow- plow phase, at which point radiative losses are becoming important to the dynamics of the remnant.

Massive stars are known for significantly altering their immediate surroundings out to several tens of parsec through their winds. 
Their ionizing radiation and fast main sequence winds will carve out a high pressure and high temperature ``bubble'', while post- main sequence (e.g. RSG) winds create dense,
$\rho \propto r^{-2}$ shells of material around the star \citep[e.g.][]{Dwarkadas05}. Nearby, neighboring massive stars might extend this low density bubble through their respective fast main sequence winds. 
As a first step towards more realistic environments around our progenitor stars, we considered a parameterized RSG wind profile assuming a wind with mass loss rate of $\dot{M}=2\times10^{-5}$ \msol{} $\rm{yr}^{-1}$ and a wind velocity of \kms{20}. We assume a spherically symmetric wind in order to assess main differences between different functional forms of the density distribution. 

\subsection{Hydrodynamic Calculations}

In order to study the mixing of ejecta material with the ISM following the explosion as a SN, we calculate the SN explosion of two 
type II progenitors expanding into different ambient media. 
We use the 3- dimensional Lagrangian hydrodynamics code SNSPH \citep{snsph} to model the explosion of the progenitors. 
SNSPH is a particle-based algorithm and is based on the version of SPH developed by \citet{Benz84,Benz88,Benz89}. 
The code is designed for fast traversal on parallel systems and for many architectures. The sizes (scale lengths) of the SPH particles is variable, and the time stepping is adaptive.

To strike a workable balance between resolution and consumed resources per simulation, we use a resolution of 1 million SPH particles for the progenitor, and 10 million SPH particles for the ambient medium in the calculations. 
Keeping the resolution in mass per SPH particle roughly the same between the ejecta and the surrounding material, this means we can follow the SN explosion until the debris has swept up about 10 times its ejecta mass. 
At this point, SNRs are generally considered to be in the Sedov-Taylor (ST) phase, approaching the snow- plow phase, in their evolution \citep{Gull73a}. 
The beginning of the snow- plow phase in SNRs is marked by shocks that become radiative, that is energy lost in radiation is becoming important for the hydrodynamical evolution and can not be neglected any longer in simulations.
Furthermore, if a pulsar was born in the explosion it will have established a pulsar wind nebula around it with which the reverse shock will eventually interact.
Since the nature of the compact central object is not modeled (aside from its gravitational effect), the creation of a pulsar wind nebula and the later interaction with the reverse shock is currently not tracked in these simulations. This interaction, however, does not occur until the forward shock has swept through several times the ejecta mass in surrounding material. 
Carrying the simulations out up to this stage allows for an accurate comparison with observational data of SNRs.

There is an intrinsic scatter in density and pressure in SPH methods, due to the variability of and dependence on the smoothing length.
In these simulations, this scatter has a 1$\sigma$ error of $\sim5-10\%$ in the lowest resolution simulations. 
%\nts{(WHat is the scatter in the 50M run?)} \notes{(It looks to be the same, I think this scatter is intrinsic to the WVT setup and not sensitive to the number of particles)}
It is likely that convection in burning shells before/during stellar collapse produces density perturbations at a $\sim~10\%$ in any case \citep{ArnettM11}, so this artificial scatter should be comparable to the true initial conditions \citep{snsph}.

The dynamics of the explosion and the young SNR evolution are dominated by the hydrodynamics. Radiative losses are assumed to not become significant until the SNR enters the snow- plow phase. Nuclear burning has a negligible effect on the dynamics of the explosion. But if the explosion produced a lot of $^{56}$Ni, its decay to $^{56}$Fe can create a Ni- bubble in the inner parts of the ejecta. 
SNSPH was augmented with a nuclear reaction network code running in step with the 
SPH calculation.
The nuclear burning code consists of a 20- isotope library comprised of mostly alpha-chain reactions to track energy generation, and is capable of burning in normal and nuclear statistical equilibrium (NSE) conditions during the explosion, and following radioactive decay only for the evolution after the explosion.
Abundance tracking for this routine was achieved by adding the abundance 
information of 20 isotopes (those used in the network) to the quantities tracked per SPH particle. 
The abundances of these isotopes were initialized to the abundance profile calculated in the progenitor models at the time of core collapse. The chemical evolution of each 
progenitor was calculated with TYCHO \citep{YoungA05} using a 177 isotope network that is complete through the Fe- peak. The 177 isotopes from the progenitor models were condensed to 20 isotopes by rebinning and renormalizing the mass fractions of the isotopes, and by adjusting the $^{56}$Fe abundance to keep the electron fraction ($Y_e$) consistent.
These abundances were followed in the code along with each particle, but chemical 
diffusion was neglected. The only physical effect that influenced the chemical composition of an SPH 
particle was through the nuclear burning or radioactive decay calculated by the network. Details on the reaction network 
can be found in \citet{eyfr11} and \citet{YoungA05}.

The size of the network is sufficient to calculate the energy generation accurately, but it is not sufficient for the accurate calculation of the abundances of the tracked elements. Thus abundances are given with the understanding that they have predictive power for {\it qualitative} behavior only. To calculate accurate abundances, the 177 isotopes from the progenitor models and the thermodynamic tracks calculated in SNSPH will be fed in to the Burnf code in a post- processing step \citep[see][for details]{YoungF07}. The calculation of detailed abundance distribution is planned for a follow-up publication.

%~~~~~~~~~~~~~~~~~~~~~~~~~~~~~~~~~~~~~~~~~~~~~~~~~~~~~~~~~~~~
%%%%%%%%%%%%%%%%   SECTION 3   %%%%%%%%%%%%%%%%
%~~~~~~~~~~~~~~~~~~~~~~~~~~~~~~~~~~~~~~~~~~~~~~~~~~~~~~~~~~~~
\section{Results\label{s:results}}

\begin{figure}[bth]
  \centering
  \includegraphics[width=0.69\textwidth]{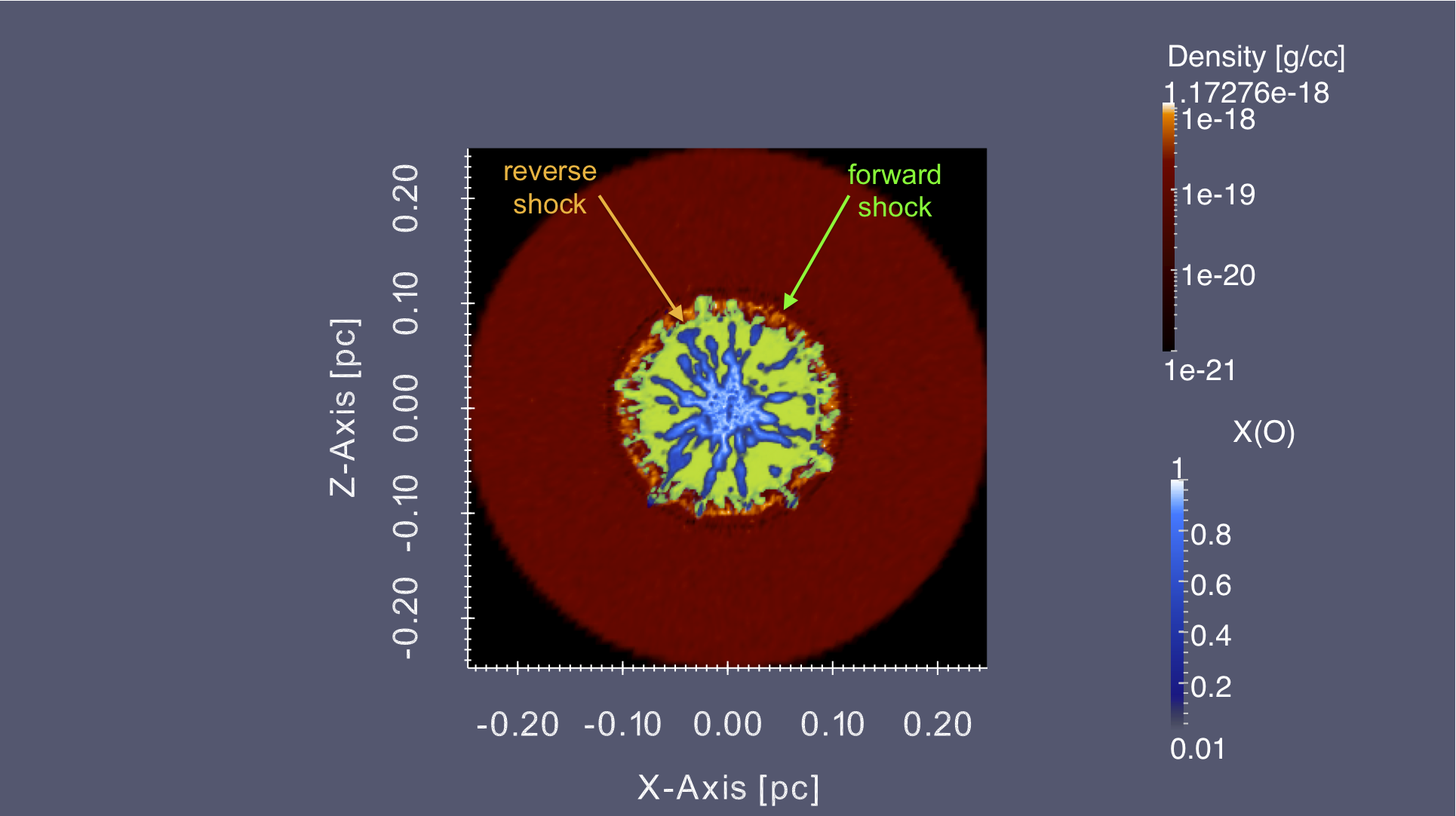}
  \includegraphics[width=0.69\textwidth]{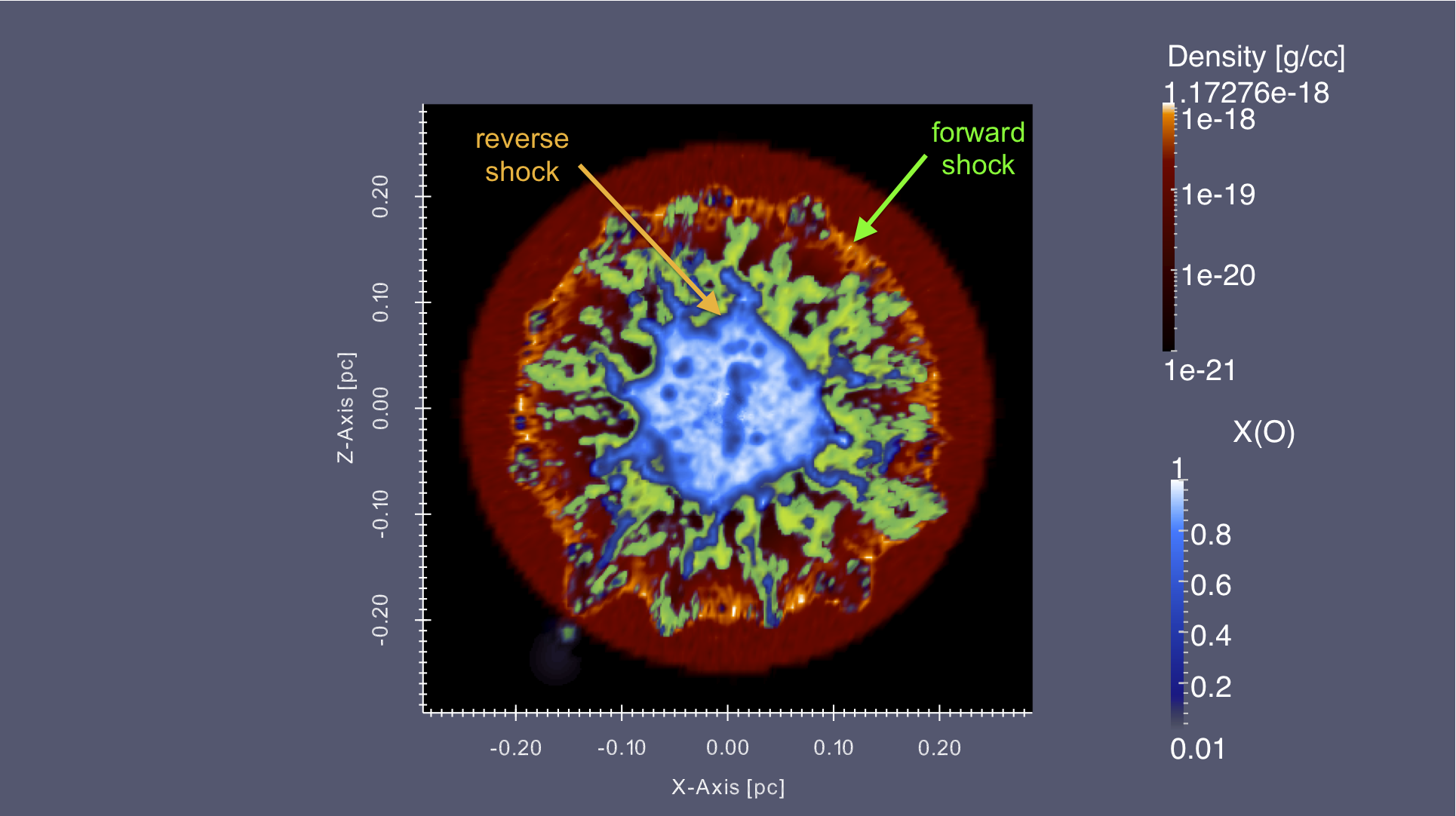}
  \includegraphics[width=0.69\textwidth]{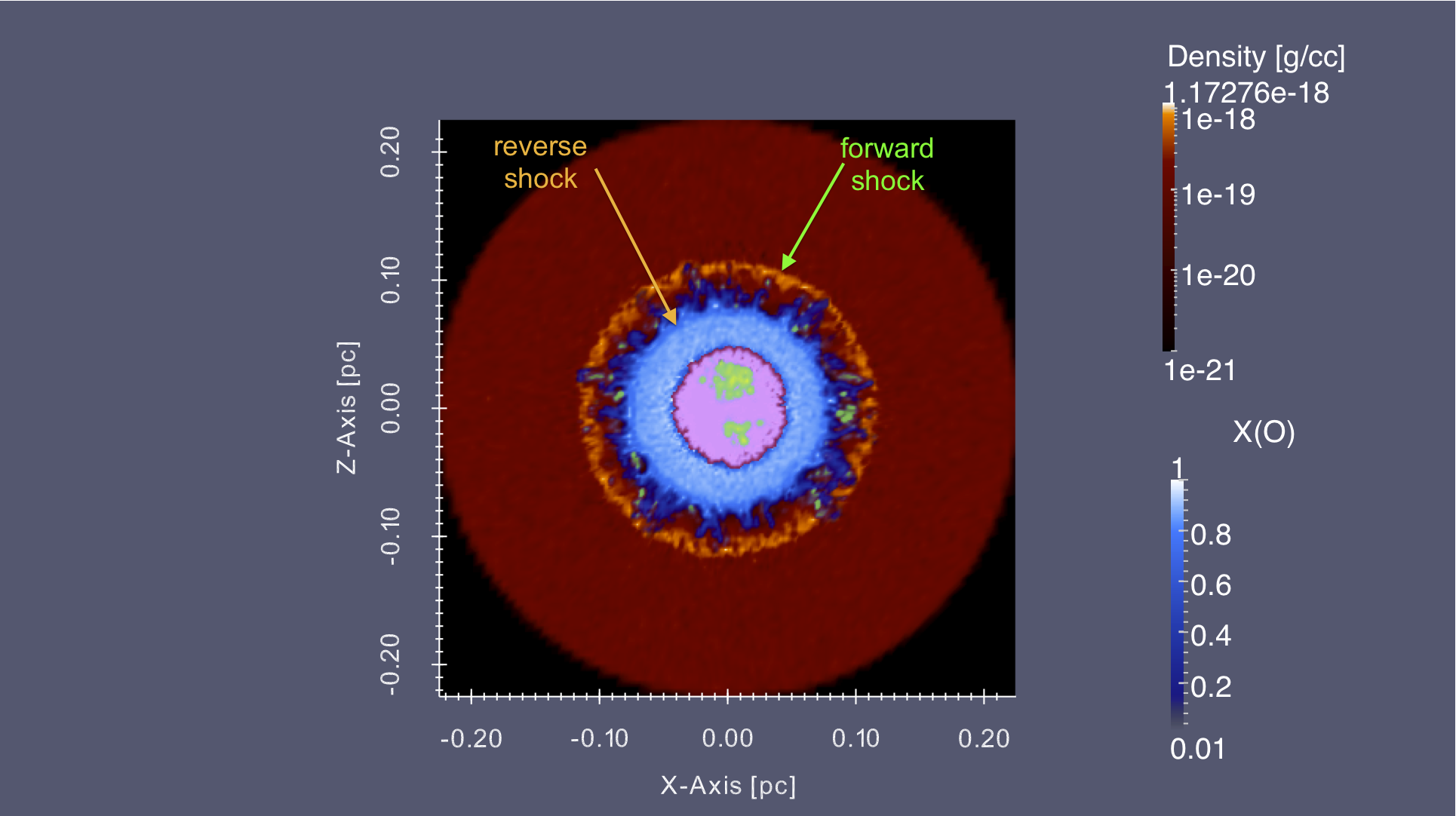}
  \caption{
  Cross sectional color gradient maps at two different times of the simulations of two of the \msol{20} SN explosions into a molecular cloud -type surrounding medium. The top panel shows the morphology of the remnant when the mass ratio m$_{ej}/$m$_{sweptup} \sim 1$, the middle panel when it is $\sim 8$. The Red color gradient indicates the
density, and the blue gradient indicates the stellar O -rich material. The stellar H- rich material was colored in green, and the stellar Fe- rich material (where present) in magenta for reference. The location of the forward and reverse shocks are indicated in each plot. 
For comparison, the bottom panel shows the \msol{16} SN when the mass ratio m$_{ej}/$m$_{sweptup} \sim 6$. Note that the forward shock is nearly undistorted. Since the \msol{16} binary progenitor retained very little of its H -envelope, no strong RT -instabilities developed in the explosion as in the \msol{20} (viz. there are no blue filaments interior to the reverse shock). The presence of these instabilities influence the degree of distortion of the forward shock in the later stages of the SNR expansion. Also note, the wind environment that would be present around physical stars was ignored.}
  \label{fig:SNRsims}
\end{figure}

\subsection{The Remnant- RT instability}
We find that a reverse shock is launched when $\rm{m_{CSE}}/\rm{m_{ej}} \simeq 0.01$. At the same time the contact discontinuity shows high mode distortions from sphericity. At this point, a dense shell of swept up ambient medium has piled up in front of the expanding ejecta, and has begun to slow it down. The outermost ejecta material that is thus slowed down in turn results in a dense shell of ejecta gas. Initially, these two dense shells are separated by a contact discontinuity. The shell of decelerated ejecta material then travels inwards (in mass coordinate) through the ejecta as a reverse shock.
The remnant reverse shock decelerates the expanding ejecta material, that is a gas of a lower density (the expanding ejecta) has a net acceleration towards a gas of a higher density (the reverse shock). A higher entropy fluid (the undecelerated, expanding ejecta) experiencing a net acceleration towards a lower entropy fluid (the decelerated ejecta in the reverse shock) is hydrodynamically unstable to an RT instability. 

The initial distortions at the contact discontinuity (CD) continue to grow in the onset of the RT instability into approximately radially outward growing filaments. Since this RT instability occurs principally by the same process as the RT instability in the star as it explodes, we will refer to the former as the ``remnant- RT instability'', and the latter as the ``SN- RT instability''. The remnant-  RT filaments and plumes mix ejecta material upstream into the circumstellar environment (CSE) and ambient material downstream into the ejecta. It should be noted that the term ``ejecta'' here refers to all of the stellar debris ejected in the explosion, i.e. includes the H- rich and He- rich shells of the progenitor star that were present at the time of explosion. This mixing smears out the CD fairly fast, so that it becomes difficult to describe a well defined boundary between the ejecta and the surrounding material. The remnant- RT instability continues until the reverse shock reaches the center of the expansion and then ceases. The reverse shock was only observed to reach the center and disappear in the \msol{16} + CSE runs, after the forward shock had swept up about 20 times the ejecta mass in CSE material. In all other scenarios the forward shock reached the edge of the simulation domain before the reverse shock had time to reach the center. 

The \msol{20} progenitor develops a strong SN- RT instability in the explosion itself, which causes plumes of material from the oxygen- rich layer to grow out to almost the outer edge of the hydrogen envelope. The clumps at the tip of the RT plumes are around a factor of 30 overdense with respect to the immediately surrounding gas. It was found in \citep{eyfr11} that this factor is weakly dependent on the resolution of the simulations, thus the true ratio is expected to be slightly higher.  
The SN- RT fingers then influence some RT fingers in the ejecta- ISM interaction. From a visual inspection, a strong but not unique correlation is observed between the position of the overdense SN- RT fingers and the sites/nodes where remnant- RT fingers form (see Figure \ref{fig:SNRsims}). Some remnant- RT filaments arise that appear to have no link to a SN- RT finger, and are thus also not enriched in metal- rich ejecta (instead show a near normal composition). It is also observed that in some cases, multiple remnant- RT filaments visually appear to grow from a single SN- RT filament. 

\begin{figure}
  \centering
  \includegraphics[width=1.\textwidth]{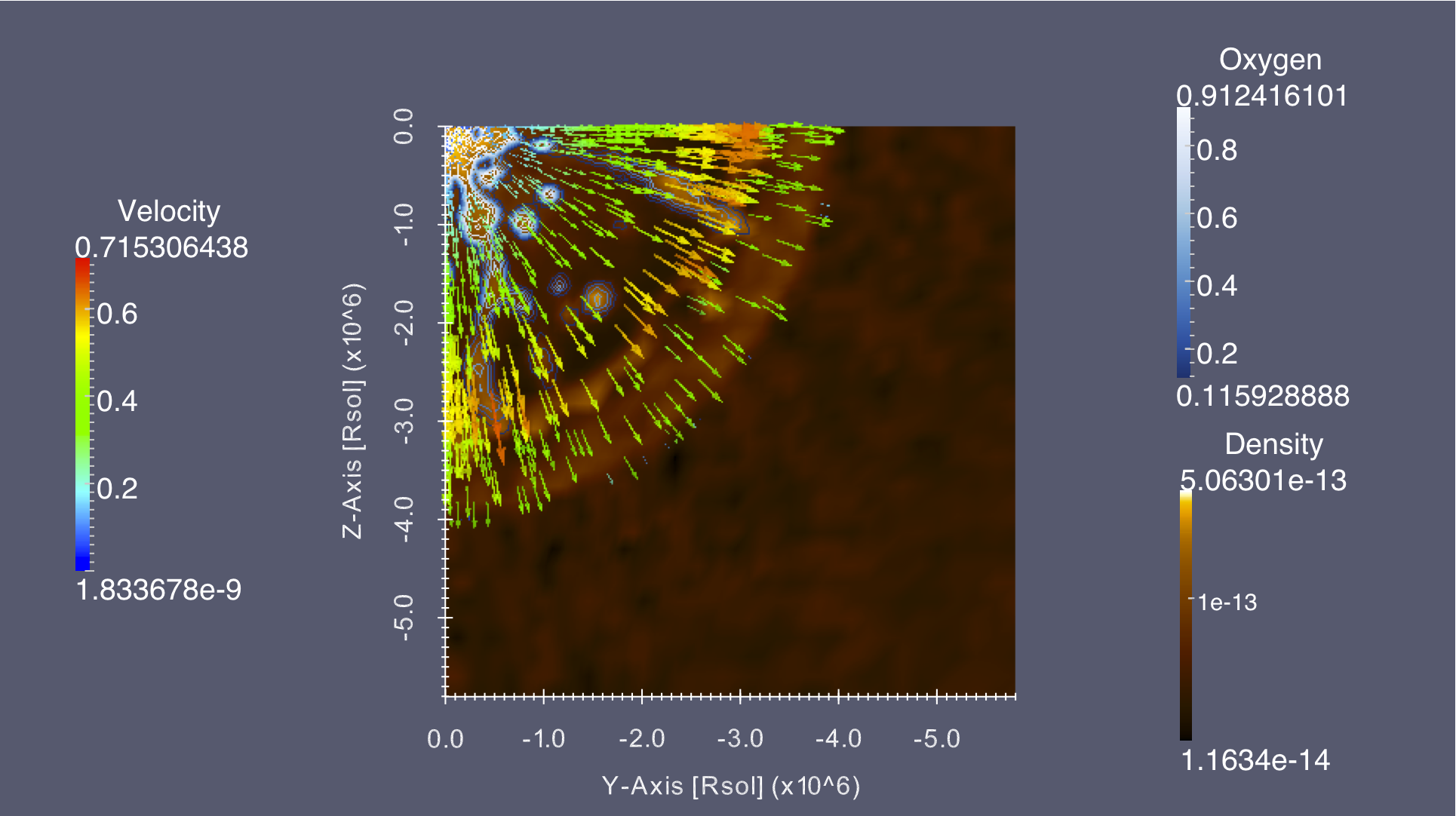}
  \caption{This figure shows a close- up of a slice through the \msol{20}+MC SNR at a time when 
  the remnant reverse shock has just reached the tip of the longest SN- RT filaments. The reverse shock and the CD are at a distance of \rsol{$\sim3\times10^6$}, the forward shock at  \rsol{$\sim4\times10^6$}. Apparent is the onset of the 
  remnant- RT instability. A cyan SN- RT finger is apparent near the abscissa that has just reached 
  the remnant- RT instability and shows the fastest velocity near its tip. }
  \label{fig:remnantRT}
\end{figure}

The remnant RT- instability arises independently of the SN- RT fingers initially, since it forms in a region in the ejecta that has not been modified by the SN-RT instability. Thus no correlation between the two would be expected in the eventual morphology. At the point when the remnant RT instability is set up, the SN- RT clumps expand homologously with the ejecta, i.e. the velocity is proportional to the radial distance from the center only. The clumps are overdense by about 1.5 orders of magnitude, however, meaning they have a higher momentum than the adjacent gas (at the same radius). That is, they experience a smaller degree of deceleration from the impact of the reverse shock as it traverses them. They thus are able to ``catch up'' to the CD where the remnant- RT instability is forming (see Figure \ref{fig:remnantRT} for a visual), and do so fairly quickly, after the RT instability has been set up, but before it becomes nonlinear and the distinctive RT fingers become apparent. This process in the remnant appears to be taking on the order of years to decades. 

The initial mode of the remnant RT instability is high, with a length scale on the order of a couple of smoothing lengths. It is manifest in a velocity distribution showing the typical characteristics of RT- induced convective flows as joint convective cells that form a web- like pattern at the CD, where the walls of the convective cells, the convective downflows, have a higher density than the centers where the upflows of the convection occurs. That is the 'web', or higher density perturbations, cover a larger area than the 'cells', or lower density spots. Conversely, the SN-RT clumps have a higher probability of coinciding with a peak in the density web, into which it then moves and eventually causes one or more remnant- RT fingers to grow. 

There are occasions where a SN-RT clump coincides with a density- trough (i.e. with a convective cell center). Since they go through a low density region in the reverse shock/ CD structure, they are moving through less material and are thus decelerated less. They retain a higher growth rate and are able to grow to larger radii, and eventually punch through the forward shock.

The remnant- filaments into which the SN- filaments grew transport metal- rich ejecta material out to the outer edge of the remnant. In the continued growth some punch through the SN blast wave moving into the ISM and distort it. The protruding filaments move supersonically into the ISM material and create bow shock structures around each of them. This creates a number of dimples and bumps in the surface that describes the forward shock. In the \msol{20} + MC run, these protrusions extend to distances about 20\% greater than the nominal forward shock. These bow- shock structures are of lower density than the gas in the forward shock. 

Some material at the tip of the remnant- filaments is ablated by and entrained into the bow shock flow which thus mixes metal- rich SN ejecta into the space between the filaments. The material between the filaments is therefore not pure ISM material, but ISM material that has been slightly enriched with SN ejecta. 

The \msol{16} did not experience a SN- RT instability in the explosion since it was lacking a hydrogen envelope. Consequently, the surface of the forward shock remains close to spherical, and the O- rich material that is mixed out remains a further distance behind the forward shock (see Figure \ref{fig:SNRsims}). 

\subsection{Ni- bubble Considerations}
The \msol{16} explosion resulted in the formation of a neutron star and in the ejection of \msol{$\sim 0.1$} of \Ni~ which is distributed near the center in approximately spherically symmetric fashion. \Ni~ decays to \Fe~ via the emission of two gamma rays which have a high chance of being trapped close to their production site and heat this material. Assuming all of the decay energy is trapped, a Ni- bubble is observed to form in the continued expansion of this explosion, which also creates a thin shell of the Si- rich material around it. This shell is eventually destroyed by the remnant- RT fingers as they reach down, break the symmetry of the shell, and mix Si and Fe up into the inner part of the O- rich material. Although some material was mixed across the Si- Fe interface, the majority of the Si remained at a larger radius from the explosion center than the Fe. 

Despite the relatively high explosions energies, the \msol{20} explosion experienced enough fallback to create a low- mass black hole. As a corollary, any Fe- group elements that were produced in the explosions were accreted into the black hole instead of being ejected, thus no Ni- bubble formed after the explosions of those progenitors. Ni could be ejected, though, if the explosion had been asymmetric (e.g. bipolar).

The fall back of material onto the proto- neutron star was tracked in all explosion simulations. In the \msol{16} explosion, the dynamics of the fall back resulted in 1- 2 large RT- plumes in the accreting material, which produced a slightly overdense lobe of \Ni- rich material that was shifted slightly from the center. This distribution did not result in a noticeable departure from spherical symmetry of the Si- shell. This slight concentration of Fe- rich material was maintained after the reverse shock passed this material.
 
\subsection{RSG wind profile}
The simulation of the explosion expanding into an RSG wind profile in the cold neutral medium develops two remnant reverse shocks. First, a reverse shock is set up as the ejecta expands into the RSG. 
%\notes{(I don't know if this would heat the ejecta to an X- ray emitting state)}. 
At the time this reverse shock is set up, the explosion reverse shock is still traversing the central parts of the ejecta. The main effect of the RSG reverse shock is to clump the RT fingers from the explosion into more overdense and narrower fingers, and cause their growth into the RSG material. 

Once the forward shock has traversed the RSG material and moved into the cold neutral medium, a second remnant- reverse shock is set up at the interface between the RSG and ISM material, which is RT unstable as well. This sets up filaments of RSG material mixing into the ISM material, which creates the typical filamentary SNR structure. However, this mixes two materials of approximately solar composition, that is, the emission spectra of this interaction region would be that of solar composition material. The metal- rich ejecta is not reached until near the end of the simulation, when the remnant has grown to a significant extend (to the edge of the simulation space), and the second remnant reverse shock has propagated far enough inward to reach metal- rich ejecta. It can be noted again, though, that not all ejecta filaments correspond to metal- rich ejecta material. Furthermore, there is no strong correlation between the filaments in the RSG material and those in the ejecta material. 

\begin{figure}[bth]
  \centering
  \includegraphics[width=0.8\textwidth]{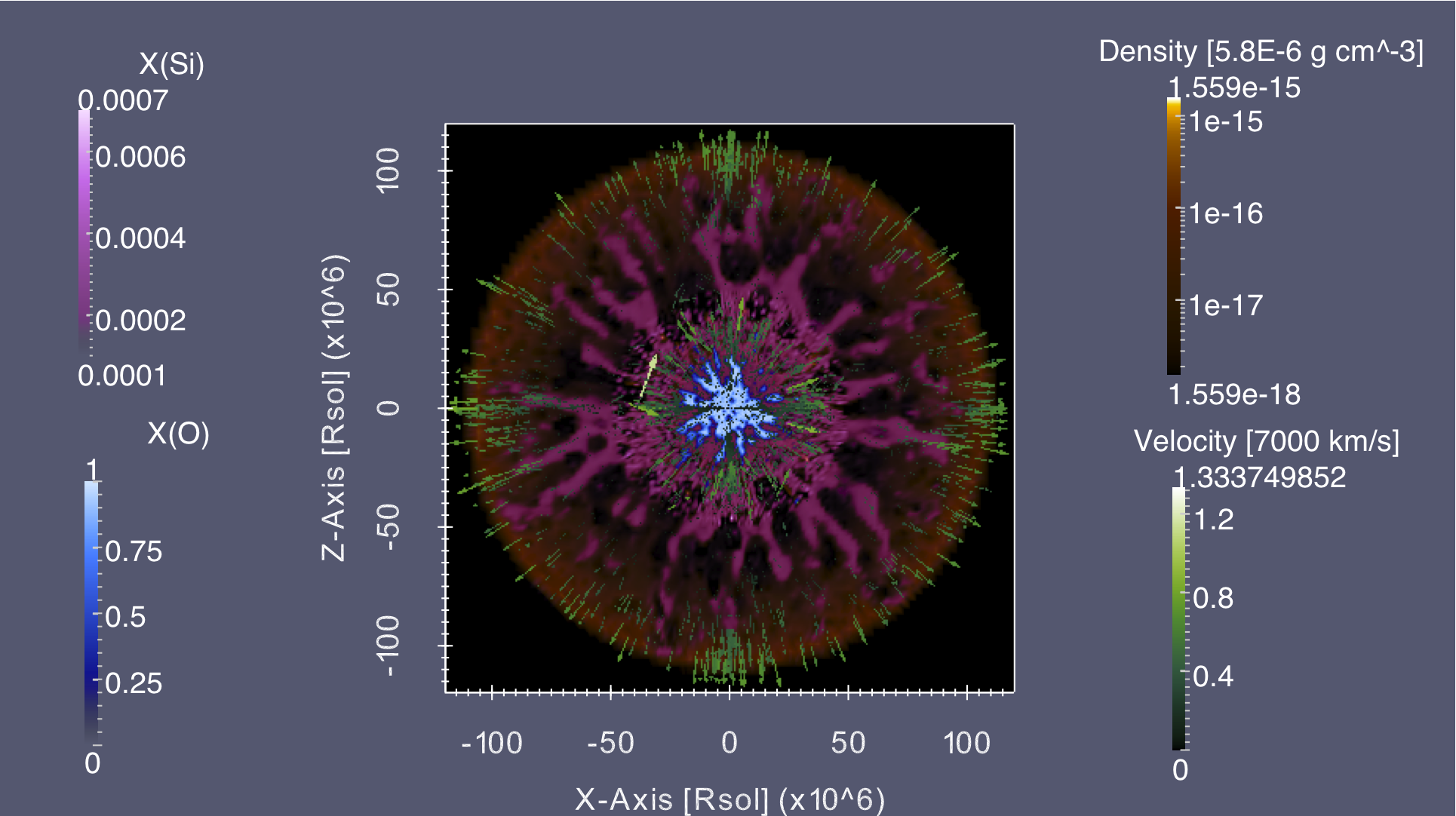}
  \caption{Shown is a cross section through the center of the \msol{20} SN + RSG wind + cold neutral medium at the end of the simulation. The velocity (given in 7000 km/s) at randomly chosen locations are shown by the green arrows (the apparent concentration of arrows near the axes is due to the regridding of the SPH distribution). The (second) remnant reverse shock can be seen at a radius of \rsol{$\sim2.5\times10^7$}, the forward shock is approaching the outer boundary of the simulation space. Remnant- RT filaments of RSG wind material (visible as the red/magenta structures) extend from the reverse shock to the forward shock and are distorting it in a few places (e.g. the clump at $(x=7\times10^7,~z=-8\times10^7)$). They are about 0.5 of an order of magnitude overdense with respect to the ISM material that they are mixing into. O- rich ejecta is shown in blue, and the Si abundance in the ISM and wind material in magenta, to emphasize the filaments formed by the (second) remnant reverse shock. }
  \label{fig:RSGdist}
\end{figure}

Figure \ref{fig:RSGdist} demonstrates the remnant structure that is created in this particular SN+RSG+CNM setup. It can clearly be seen that the RSG filaments (in magenta hues) are distinct from the ejecta filaments (viz. the O- rich filaments in blue), and that an extensive interaction region was created that contains almost no metal- rich material.

\clearpage

%\begin{table}[bth]
%\centering
%\begin{tabular}{lccc}
%\hline \hline \\
%  Simulation & m$_{ej}$/m$_{swept-up}$ & Simulation age (yr) & K. E. (\foe{})
%\\ \hline \\
%\multirow{3}{*}{
%\msol{16}+CN} & 0.10& 11.7 & 6.6
%\\
%& 0.84 & 43 & 4.4
%\\
%& 5.1 & 134 & 2.5
%\\
%\\
%\multirow{4}{*}{\msol{16}+MC} & 0.09& 1.0 & 6.7
%\\
% & 1.0 & 4.4 & 4.3
%\\
% & 5.0 & 12 & 2.3
%\\
% & 10.7 & 22 & 1.9
%\\
%\\
%\multirow{4}{*}{\msol{20}+MC} &0.11 & 5.5 & 1.8
%\\
% & 1.0 & 18 & 1.2
%\\
% & 4.9 & 48 & 0.74
%\\
% & 9.5 & 76 & 0.6
%\\ 
%\\
%\multirow{4}{*}{\msol{20}+RSG+CN} &0.24 & 0.002 & 1.2
%\\
% & 1.2 & 2.8 & 45
%\\
% & 4.7 & 79 & 23
%\\
% & 9.6 & 153 & 20
%\\ 
%\\
%\multirow{3}{*}{\sout{\msol{25}+disk+CN}} &0.12 & 35 & 6.2
%\\
% & 0.94 & 121 & 4.4
%\\
% & 5.3 & 342 & 2.7
%\\
%\\
%\multirow{3}{*}{\sout{\msol{25}-j4+MC}} &0.11 & 39 & 5.9
%\\
% & 0.95 & 128 & 3.9
%\\
% & 4.9 & 338 & 2.5
%\\
%\\ \hline \\
%\end{tabular}
%\caption{Kinematic Properties}
%\label{tb:kin}
%\end{table}
%

%~~~~~~~~~~~~~~~~~~~~~~~~~~~~~~~~~~~~~~~~~~~~~~~~~~~~~~~~~~~~
%%%%%%%%%%%%%%%%   SECTION 4   %%%%%%%%%%%%%%%%
%~~~~~~~~~~~~~~~~~~~~~~~~~~~~~~~~~~~~~~~~~~~~~~~~~~~~~~~~~~~~
%\input{./discuss.tex}

%~~~~~~~~~~~~~~~~~~~~~~~~~~~~~~~~~~~~~~~~~~~~~~~~~~~~~~~~~~~~
%%%%%%%%%%%%%%%%   SECTION 5   %%%%%%%%%%%%%%%%
%~~~~~~~~~~~~~~~~~~~~~~~~~~~~~~~~~~~~~~~~~~~~~~~~~~~~~~~~~~~~
\section{Discussion\label{s:conclude}}

In this paper, we present first simulations investigating the
3-dimensional morphology created in the SN ejecta from the launch of
the explosion through the formation of a young SNR.  Although the
growth of RT instabilities is well-studied in the
literature \citep[for reviews, see][]{Glimm_ea01,Dimonte_ea04}, some
discrepancies persist between the results of different groups.  In
part, this is due to different numerical schemes that damp out or
modify the evolution of the high wavenumber perturbations
\citep{Glimm_ea01}.  But the growth of RT instabilities
is also very sensitive to the spectrum of the initial broadband
perturbations \citep{CookD01}, and the mode-coupling between low- and
high-mode instabilities can significantly alter the evolution of the
RT instabilities
\citep{Calder_ea02,Dimonte_ea04,Woodward_ea10}.  Although the remnant
instabilities initially grow independently, as the SN-RT instabilities
catch up to the forward shock, these instabilities can shape the
remnant and interact with the remnant instabilities.  As we expect
from these studies of RT instabilities, the final remnant
structure looks very different from remnant studies focusing only on
remnant-RT.  The SN-RT instabilities are a key aspect in understanding
the turbulent structure in remnants.

For many of the broad features produced in our simulations (and the
observed in remnants), we found that 11 million SPH particles provide
an adequate resolution to study that problem (though we are far from
fully resolving the instabilities).  Comparison between a 1 million
and a 50 million particle simulation of the SN explosion only
\citep{eyfr11} has shown that at 1 million particles (i.e. the same
resolution used for the progenitors in this study) the RT instability
is not fully resolved, though enough to give useful information, and
that shear flow, or Kelvin- Helmholtz (KH), instabilities are likely
strongly suppressed. This is an artifact of the current implementation
of the artificial viscosity in the SNSPH code which is being
resolved. If one assumes that AMR codes do accurately resolve KH
instabilities, we can make the prediction that the RT filaments should
be slightly more smeared out and broken up into branches. However, if
magnetic fields are present in the gas, these would suppress the
shear- flow mixing across the ``surface'' of the RT filaments, making
them less smeared out. We further expect that a much higher resolution
(e.g. 1 billion SPH particles) for the simulations might eventually
yield a higher RT mode, i.e. more numerous and narrower RT
filaments. However, all these considerations are expected to provide
mostly quantitative refinements to the properties of the features that
are created. The qualitative picture that we find from these
simulations is expected to remain the same.

In this paper we only considered spherically symmetric cases of both
the SNe and the surrounding media.  Since we started the 3D aspect of
the SN explosions in this study after the bounce shock has been
revived, we effectively ignored any asymmetries that might have arisen
from the explosion engine (i.e. during the revival of the shock), or
that were present in the material before the explosion. The explosion
engine itself is believed to operate through instabilities like
Neutrino Driven Convection or the Standing Accretion Shock
Instability, which results in partial turnover of the innermost
material (i.e. of the Fe- rich and Si/S- rich material). Furthermore,
\citep{ArnettM11} have recently shown that the structure of a massive
star immediately before the onset of core collapse is far from
symmetric or onion-skin layered. Instead, the dynamics of the Si- and
O- burning shells lead to vigorous burning and overturning of
material, which cause large deviations in the compositional and
density structure and energy generation rate. Because of the time
scale associated with this turbulence, the explosion of the star is
very likely to occur in this highly distorted state of the inner
burning shells. Both of these processes should leave global imprints
on the distribution of the chemical elements. These global imprints
would be retained even in the remnant- RT instability (though small
scale variations would be destroyed).

The typical SNR structure is set up once the swept up mass becomes comparable to the ejecta mass. 
At this point, a reverse shock is moving inwards (in mass coordinate) through the ejecta which heats and ionizes it, thus making it observable in X- rays. Predictions for the precise nature of the emission requires the detailed calculations of the ionization state and history, which is beyond the scope of this paper. However, a qualitative assessment of the emission properties can nevertheless give useful insight. 

As noted before, in this paper the term ``ejecta'' refers to all of the ejected stellar debris, whereas we use the term ``metal- rich ejecta'' to explicitly denote the part of the stellar debris originating from the O- rich and deeper shells of the star. 
As the reverse shock traverses the ejecta gas, it compresses and heats it to X- ray emitting temperatures. The density of the gas is low enough that its ionization occurs collisionlessly; the actual mechanism of this is not well understood though. The X-ray intensity per wavelength of a parcel of this gas is a function of its ionization state, its temperature, its chemical composition, and its density. Generally, higher density or more metal rich gas emits more brightly, and an overabundance of a certain element is reflected in a brighter emission line. The remnant- RT filaments, which consist of metal- rich ejecta gas, are therefore expected to emit more brightly than the gas in between them, which consists of ISM- like material (which might have been slightly enriched with metals from the ejecta). The remnant- RT instability, which causes overdensities and large scale turnover of gases of different composition, thus has a direct effect on the local emission properties of SNRs. 

In all simulations the remnant reverse shock over time slowly causes the interpenetration of ISM or RSG material with the ejecta material, though the two gases do not fully mix to a homogenous mixture before the end of the simulations. 
The specific internal energy in the interaction region varies by about one order of magnitude, with the ejecta filaments tending towards lower specific internal energies. If the temperature is purely due to the material pressure, it should then vary by an order of magnitude across this region as well (neglecting any ionization effects). 
In the simulations of the RSG progenitors expanding into different CSEs this might be important to note, since the onset of the remnant- RT instability occurs in the interaction between the ejected H- envelope and the CSE. Both materials have compositions that are approximately solar (cosmic) abundances. That is, the X- ray emission of this interaction might show filamentary features corresponding to the remnant- RT filaments, but would show spectra of normal composition (though likely small ionization time scales due to the very recent reverse shock passage).

\acknowledgements
The authors graciously acknowledge the use of the Texas Advanced Computing Center (TACC) for the hydrodynamics calculations. This work has been supported in part by SAO under Chandra grants GO0-11075A and GO0-11076X.

%}

\bibliographystyle{../natbib/astron.bst}
\bibliography{SNRinteractions,emission,explosion,StelEvol,CasA,g292,mypapers,hydro}

\end{document}